\documentclass[12pt]{article}
\usepackage{amsfonts}

\usepackage{amsfonts}
\usepackage{amssymb}
\usepackage{amscd}

\def\D{\hbox{D\kern-.73em\raise.25ex\hbox{-}\raise-.25ex\hbox{ }}}
 \def\d{\hbox{d\kern-.33em\raise.75ex\hbox{-}\raise-.75ex\hbox{}}}

\def\GGG{\frak G }
\def\gr3{\GGG\,(\SSS_3)}

\def\gr2{\GGG\,(\SSS_2)}

\def\SSS{\frak S}

\def\ed{\end{document}}
\def\beq{\begin{equation}}
\def\eeq{\end{equation}}
\def\bea{\begin{eqnarray}}
\def\eea{\end{eqnarray}}
\def\ba{\begin{array}}
\def\ea{\end{array}}
\def\bi{\begin{itemize}}
\def\ei{\end{itemize}}

\newcommand{\bp}{\noindent\begin{minipage}[c]}
\newcommand{\ep}{\end{minipage}}

\date{}

\begin{document}

 \title{\Large\bf  $p$-Adic Degeneracy of the Genetic Code}

\author{Branko
Dragovich$^a$\,\footnote{\textsf{\, E-mail:\,dragovich@phy.bg.ac.yu}}\,\, and Alexandra Dragovich$^b$\\
 {\it $^a$Institute of Physics, P.O. Box 57} \\ {\it 11001
Belgrade, Serbia} \\ {\it $^b$Vavilov Institute of General
Genetics}\\ {\it Gubkin St. 3, \, 119991  Moscow, \,Russia}}

\maketitle

\begin{abstract}
Degeneracy of the genetic code is a biological way to minimize
effects of the undesirable  mutation changes. Degeneration  has a
natural description on the $5$-adic space of $64$ codons
$\mathcal{C}_5 (64) = \{ n_0 + n_1 \, 5 + n_2\, 5^2 \,:\,\, n_i = 1,
2, 3, 4 \}\,,$ where $n_i$ are digits related to nucleotides as
follows:  C = 1, A = 2, T = U = 3, G = 4. The smallest $5$-adic
distance between codons joins them into $16$ quadruplets, which
under $2$-adic distance decay into $32$ doublets. $p$-Adically close
codons   are assigned to one of $20$ amino acids, which are building
blocks of proteins, or code termination of protein synthesis. We
shown that genetic code multiplets are made of the $p$-adic  nearest
codons.
\end{abstract}

\bigskip

\section{Introduction}

Genetic information in living systems is contained in the
desoxyribonucleic acid (DNA) sequence. The DNA macromolecules are
composed of two polynucleotide chains  with a double-helical
structure. The building blocks of the genetic information  are four
nucleotides called: adenine (A), guanine (G), cytosine (C) and
thymine (T). A and G are purines, while C and T are pyrimidines.
Nucleotides are arranged along  double helix through base pairs A-T
and C-G. The DNA is packaged in chromosomes which are localized in
the nucleus of the eukaryotic cells. One of the basic processes
within DNA is its replication. The passage of DNA gene information
to proteins, called gene expression, performs by the messenger
ribonucleic acids (mRNA), which are usually  single polynucleotide
chains. The mRNA are synthesized in the first part of this process,
known as transcription, when nucleotides A, G, C, T from DNA are
respectively transcribed into their complements U, C, G, A of mRNA,
where T is replaced by U (U is the uracil). The next step is
translation, when the information coded by codons in the mRNA  is
translated into proteins. In this process two other RNA's are
involved: transfer tRNA and ribosomal rRNA. Codons are ordered
sequences of three nucleotides taken of the A, G, C, U. Protein
synthesis in all eukaryotic cells performs in the ribosomes of the
cytoplasm.

The genetic code relates the information of the sequence of codons
in mRNA to the sequence of amino acids in a protein. Although there
are about dozen codes (see, e.g. \cite{sorba1}), the most important
are two of them: the eukaryotic code and the vertebral mitochondrial
code. In the sequel we shall mainly consider the vertebral
mitochondrial code, because it looks the simplest one and the others
can be regarded as its modifications. It is obvious that there are
$4 \times 4\times 4 = 64$  codons. However (in the vertebral
mitochondrial code), $60$ of them are distributed on the $20$
different amino acids and $4$ make stop-codons, which serve as
termination signals. According to experimental observations, two
amino acids are coded by six codons, six amino acids by four codons,
and twelve amino acids by two codons. This property that to an amino
acid corresponds more than one codon is known as genetic code
degeneracy. This degeneracy is a very important property of the
genetic code and gives an efficient way to minimize effects of the
undesirable mutation changes. Since there is a huge number (about
$10^{80}$) of all possible assignments between codons and amino
acids,  and only a very small number (about dozen) of them is
represented in living cells, it has been a persistent theoretical
challenge to find an appropriate model explaining contemporary
genetic codes. Still there is no generally accepted explanation of
the genetic code. For a detail and comprehensive information on
molecular biology aspects of DNA, RNA and genetic code one can see
Ref. \cite{watson}. It is worth mentioning that human genome, which
presents all genetic information of the homo sapiens, is composed of
about three billions DNA base pairs and contains  more than 20.000
genes.

Modeling of DNA, RNA and genetic code is a challenge as well as an
opportunity for modern mathematical physics. An interesting model
based on the quantum algebra $\mathcal{U}_q (sl(2)\oplus sl(2))$ in
the $q \to 0$ limit was proposed as a symmetry algebra for the
genetic code (see \cite{sorba1} and references therein). In a sense
this approach mimics quark model of baryons. To describe
correspondence between codons  and amino-acids, it was constructed
an operator which acts on the space of codons and its eigenvalues
are related to amino acids. Besides some successes of this approach,
there is a problem with rather many parameters in the operator.
There are also papers \cite{hornos i drugi}  starting with
64-dimensional irreducible representation of a Lie (super)algebra
and trying to connect multiplicity of codons  with irreducible
representations of subalgebras arising in a chain of symmetry
breaking. Although interesting as an attempt to describe evolution
of the genetic code  these Lie algebra approaches did not succeed to
get its modern form.   For a very brief review of these and some
other theoretical approaches to the genetic code one can see Ref.
\cite{sorba1}.

Recently we introduced a $p$-adic approach to the DNA, RNA sequences
and genetic code \cite{dragovich1}. Let us mention that $p$-adic
models in mathematical physics have been actively considered since
1987 (see \cite{freund}, \cite{vladimirov1} for early reviews and
\cite{dragovich2}, \cite{dragovich3} for some recent reviews). It is
worth noting that $p$-adic models with pseudodifferential operators
have been successfully applied to interbasin kinetics of proteins
\cite{avetisov3}.  Some $p$-adic aspects of cognitive, psychological
and social phenomena have been also considered \cite{khrennikov1}.
The present status of application of $p$-adic numbers in physics and
related branches of sciences is reflected in the proceedings of the
2nd International Conference on $p$-Adic Mathematical Physics
\cite{proceedings}. The main goal of this paper is to present
$p$-adic root of the genetic code and, in particular, its
degeneracy.

\section{$p$-Adic space of codons}

An elementary introduction to $p$-adic numbers  can be found in the
book \cite{gouvea}. However, for our purposes we will use here only
a bit of $p$-adics, mainly a finite set of integers and ultrametric
distance between them.

Let us introduce the  set of natural numbers

\beq \mathcal{C}_5 (64) = \{ n_0 + n_1\, 5 + n_2\, 5^2 \,:\,\, n_i =
1, 2, 3, 4 \}\,,   \label{2.1}\eeq where $n_i$ are digits related to
nucleotides by the following assignment: C = 1,\, A = 2,\, T = U =
3,\, G = 4. This is an expansion to the base $ 5$. It is obvious
that $5$ is a prime number and that the set $\mathcal{C}_5 (64)$
contains $64$ numbers between $31$ and $124$ (in the usual base 10).
In the sequel we shall denote elements of $\mathcal{C}_5 (64)$ by
their digits to the base $5$ in the following way: $ n_0 + n_1\, 5 +
n_2\, 5^2 \, \equiv n_0\, n_1\, n_2$. Note that here ordering of
digits is the same as in the expansion (\ref{2.1}), i.e this
ordering is opposite to the usual one. There is now evident
one-to-one correspondence between codons in letter XYZ and number
$n_0\, n_1\, n_2$ representations.

In addition to arithmetic operations it is often important to know
also a distance between numbers. Distance can be defined by a norm.
On the set $\mathbb{Z}$ of integers  there are two kinds of
nontrivial norm: usual absolute value $|\cdot|_\infty$ and $p$-adic
absolute value $|\cdot|_p$ , where $p$ is any prime number. The
usual absolute value is well known from elementary courses of
mathematics and the corresponding distance between two  numbers $x$
and $y$ is $d_\infty (x, y) = |x-y|_\infty$.

The $p$-adic absolute value is related to the divisibility of
integers by prime numbers, and $p$-adic distance can be understood
as a measure of this divisibility for the difference of two numbers
(the more divisible, the shorter). By definition, $p$-adic norm of
an integer  $m \in \mathbb{Z}$, is $|m|_p = p^{-k}$, where $ k \in
\mathbb{N} \bigcup \{ 0\}$ is degree of divisibility of $m$ by prime
$p$ (i.e. $m = p^k\, m'\,, \,\, p\nmid m'$) and $|0|_p =0.$ This
norm is a mapping from $\mathbb{Z}$ into non-negative real numbers
and has the following properties:

(i) $|x|_p \geq 0, \,\, |x|_p =0$ if and only if $x = 0$,

(ii) $|x\, y|_p = |x|_p \,  |y|_p \,,$

(iii) $|x + y|_p \leq \, \mbox{max}\, \{ |x|_p\,, |y|_p \} \leq
|x|_p + |y|_p $ for all $x \,, y \in \mathbb{Z}$.

\noindent Because of the strong triangle inequality $|x + y|_p \leq
\, \mbox{max} \{ |x|_p\,, |y|_p \}$, $p$-adic absolute value belongs
to non-Archimedean (ultrametric) norm. One can easily conclude that
$0 \leq |m|_p \leq 1$.

$p$-Adic distance between two integers $x$ and $y$ is
\begin{equation}
d_p (x\,, y) = |x - y|_p \,.    \label{2.2}
\end{equation}
Since $p$-adic absolute value is ultrametric, the $p$-adic distance
(\ref{2.2}) is also ultrametric, i.e. it satisfies
\begin{equation}
d_p (x\,, y) \leq\, \mbox{max}\, \{ d_p (x\,, z) \,, d_p (z\,, y) \}
\leq d_p (x\,, z) + d_p (z\,, y) \,, \label{2.3}
\end{equation}
where $x, \, y$ and $z$ are any three integers.

The above introduced set $\mathcal{C}_5 (64)$ endowed by $p$-adic
distance we shall call $p$-adic space of codons. $5$-Adic distance
between two codons $a, b \in \mathcal{C}_5 (64)$ is

\beq d_5 (a,\, b) = |a_0 + a_1 \, 5 + a_2 \, 5^2 - b_0 - b_1 \, 5 -
b_2 \, 5^2 |_5 \,.   \label{2.4} \eeq When $a \neq b$ then $d_5
(a,\, b)$ may have three different values: (i) $d_5 (a,\, b) = 1$ if
$a_0 \neq b_0$, (ii) $d_5 (a,\, b) = 1/5$ if $a_0 = b_0 $ and $a_1
\neq b_1$, and (iii) $d_5 (a,\, b) = 1/5^2$ if $a_0 = b_0 \,,
\,\,a_1 = b_1$ and $a_2 \neq b_2 $. We see that the maximum $5$-adic
distance between codons is $1$ and it is equal to the maximum
$p$-adic distance on $\mathbb{Z}$. Let us also note that this
distance depends only on the first two nucleotides in the codons.
Use of $5$-adic distance between codons is a natural one to describe
information similarity between them.

In the case of standard distance $d_\infty (a,\, b) = |a_0 + a_1 \,
5 + a_2 \, 5^2 - b_0 - b_1 \, 5 - b_2 \, 5^2 |_\infty $, third
nucleotides $a_2$ and $b_2$  play more important role than those at
the second place (i.e $a_1 \, \mbox{and} \, b_1$), and nucleotides
$a_0$ and $b_0$ are of the smallest importance.

\section{$p$-Adic genetic code}

Living cells are  very complex systems composed mainly of proteins
which play various roles. These proteins are long linear chains made
of only 20 amino acids, which are the same for all living world on
the Earth. Different sequences of amino acids form different
proteins.

An intensive study of connection between ordering of nucleotides in
the DNA (and RNA) and ordering of amino acids in proteins led to the
experimental discovery of genetic code in the mid-1960s. Genetic
code is understood as a dictionary for translation of information
from the DNA (through RNA) to production of proteins by amino acids.
The information on amino acids is contained in codons. To the
sequence of codons in the RNA corresponds quite definite sequence of
amino acids in a protein, and this sequence of amino acids
determines a primary structure of the protein.

However, there is no simple theoretical understanding of genetic
coding. In particular, it is not clear why genetic code exists just
in the known way and not in many other possible ways. What is a
principle (or principles) used in establishment of a basic
(mitochondrial) code? What are properties of codons connecting them
into definite multiplets which code the same amino acid or
termination signal? These are only some of many questions whose
answers should lead us to make an appropriate theoretical model of
the genetic code.

\bigskip
\bigskip

\noindent 
\begin{tabular}{|c|c|c|c|}
 \hline \ & \ & \ & \\
 111 CCC Pro &  211 ACC Thr  & 311 UCC Ser & 411 GCC Ala  \\
 112 CCA Pro &  212 ACA Thr  & 312 UCA Ser & 412 GCA Ala  \\
 113 CCU Pro &  213 ACU Thr  & 313 UCU Ser & 413 GCU Ala  \\
 114 CCG Pro &  214 ACG Thr  & 314 UCG Ser & 414 GCG Ala  \\
 \hline \  & \  &  \ & \ \\
 121 CAC His &  221 AAC Asn  & 321 UAC Tyr & 421 GAC Asp  \\
 122 CAA Gln &  222 AAA Lys  & 322 UAA Ter & 422 GAA Glu  \\
 123 CAU His &  223 AAU Asn  & 323 UAU Tyr & 423 GAU Asp  \\
 124 CAG Gln &  224 AAG Lys  & 324 UAG Ter & 424 GAG Glu  \\
 \hline \  & \  & \  &   \\
 131 CUC Leu &  231 AUC Ile  & 331 UUC Phe & 431 GUC Val \\
 132 CUA Leu &  232 AUA Met  & 332 UUA Leu & 432 GUA Val \\
 133 CUU Leu &  233 AUU Ile  & 333 UUU Phe & 433 GUU Val \\
 134 CUG Leu &  234 AUG Met  & 334 UUG Leu & 434 GUG Val \\
 \hline \ & \   & \  &   \\
 141 CGC Arg &  241 AGC Ser  & 341 UGC Cys & 441 GGC Gly  \\
 142 CGA Arg &  242 AGA Ter  & 342 UGA Trp & 442 GGA Gly  \\
 143 CGU Arg &  243 AGU Ser  & 343 UGU Cys & 443 GGU Gly  \\
 144 CGG Arg &  244 AGG Ter  & 344 UGG Trp & 444 GGG Gly  \\
\hline
\end{tabular}

\bigskip
\centerline{Table: The vertebral mitochondrial code}
\bigskip

 Let us now look at the experimental Table of the vertebral
mitochondrial code and compare it with the above introduced
$\mathcal{C}_5 (64)$ codon space. To this end, codons are
simultaneously denoted by three digits and standard capital letters
(recall: C=1,\, A=2,\, U=3,\, G=4). The corresponding amino acids
are presented in the usual three-letter form.

First of all let us note that our Table is constructed according to
the gradual change of digits and, as a consequence, there is a
different spatial distribution of amino acids comparing to the
standard (Watson-Crick) table (see, e.g. \cite{sorba1}). Any of
these tables can be regarded as a big rectangle divided into 16
equal smaller rectangles: 8 of them are quadruplets which one-to-one
correspond to 8 amino acids, and other 8 rectangles are divided into
16 doublets coding 14 amino acids and termination (stop) codon (by
two doublets at different places). Note that 2 of 16 doublets code 2
amino acids (Ser and Leu) which are already coded by 2 quadruplets,
thus amino acids Serine and Leucine are coded by 6 codons. In our
Table quadruplets and doublets together form a figure, which is
symmetric with respect to the mid vertical line, i.e. it is
invariant under interchange $1 \longleftrightarrow 4$ and $2
\longleftrightarrow 3$ of the first digits in codons. Recall that
the DNA is symmetric under simultaneous interchange of complementary
nucleotides in its strands. In other words, the DNA is invariant
under nucleotide interchange $1 \longleftrightarrow 4$ and $2
\longleftrightarrow 3$ between strands.  All doublets in the Table
form a nice figure which looks like letter $\mathbb{T}$.

Now we can look at the Table as a representation  of the
$\mathcal{C}_5 (64)$ codon space. Namely, we observe that there are
16 quadruplets such that each of them has the same first two digits.
Hence $5$-adic distance between any two different codons inside a
quadruplet is
\begin{equation}
d_5 (a,\, b) = |a_0 + a_1 \, 5 + a_2 \, 5^2 - a_0 - a_1 \, 5 - b_2
\, 5^2 |_5 = |(a_2 - b_2) \, 5^2|_5 = 5^{-2}\,,  \label{5.1}
\end{equation}
because $a_0 = b_0$,  $a_1 = b_1$ and $|a_2 - b_2|_5 = 1$.

Since codons are composed of three nucleotides, each of which is
either a purine or a pyrimidine, it is natural to try to quantify
similarity inside purines and pyrimidines, as well as distinction
between elements from these two groups of nucleotides. Fortunately
there is a tool, which is again related to the $p$-adics, and now it
is $2$-adic distance. One can easily see that the $2$-adic distance
between pyrimidines  C and U is $1/2$ as the distance between
purines  A and G.  However $2$-adic distance between C and A or G as
well as distance between U and A or G is $1$ (i.e. maximum).

  With respect to the $2$-adic distance, the above quadruplets may be regarded
 as composed of two doublets: $a = a_0\, a_1\, 1$ and $b = a_0\, a_1\, 3$
 make the first doublet, and
 $c = a_0\, a_1\, 2$ and $d = a_0\, a_1\, 4$ form the second one. $2$-Adic
 distance between codons within each of these doublets is
 $\frac{1}{2}$, i.e.
 \begin{equation}
d_2 (a,\, b) = |(3 -1)\, 5^2|_2 =\frac{1}{2} \,, \quad  d_2 (c,\, d)
= |(4 -2)\, 5^2|_2 =\frac{1}{2} \,,   \label{5.2}
 \end{equation}
because $3-1 = 4 - 2 = 2$.

One can now look at the Table as a system of 32 doublets. Thus 64
codons are clustered by a very regular way into 32 doublets. Each of
21 subjects (20 amino acids and 1 termination operation) is coded by
one, two or three doublet. In fact, there are two, six and twelve
amino acids coded by three, two and one doublets, respectively.
Residual two doublets code termination signal.

To have a more complete picture on the genetic code it is useful to
consider possible distances between codons from different
quadruplets as well as from different doublets. Also, we introduce
distance between quadruplets or between doublets, especially when
distances between their codons have the same value. Thus $5$-adic
distance between a quadruplet and quadruplets  in the same column is
$1/5$, while such distance toward all other quadruplets is $1$.
$5$-Adic distance between doublets coincides with distance between
quadruplets, and this distance is $\frac{1}{5^2}$ when doublets are
inside the same quadruplet.

The $2$-adic distance between codons, doublets and  quadruplets  is
more complex. There are three basic cases: (1) codons differ only in
one digit, (2) codons differ in two digits, and (3) codons differ in
all three digits. In the first case, $2$-adic distance can be
$\frac{1}{2}$  or $1$ depending whether difference between digits is
$2$ or not, respectively.

Let us now look at $2$-adic distances between doublets coding
Leucine and also between doublets coding Serine. These are two cases
of amino acids coded by three doublets. Doublet consisting of codons
332 and 334 should be compared with doublet of codons 132 and 134.
The largest $2$-adic distance between them is $\frac{1}{2}$. We
again obtain maximum distance $\frac{1}{2}$ for Serine when we
compare doublets (311,\, 313) and (241,\, 243).

Other known codes may be regarded as some modifications of the
vertebral mitochondrial code (inside five quadruplets of
$\mathbb{T}$-like region and quadruplet coding Leucine). The
modification means that some codons change their meaning and code
either other amino acids or termination signal. So, in the universal
(standard, canonical) code there are the following changes: (i) 232
AUA: Met $\rightarrow$ Ile, (ii) 242 AGA and 244 AGG: Ter
$\rightarrow$ Arg, (iii) 342 UGA: Trp $\rightarrow$ Ter.

\section{Discussion and concluding remarks}

We have chosen $ p=5 $ as the base in expansion of an element of the
 $\mathcal{C}_5 (64)$ space of codons, because $5$ is the smallest prime
number which contains four nucleotides (A\,, T\,, G\,, C) in DNA, or
(A\,, U\,, G\,, C) in RNA, in the form of four different digits. At
the first glance, because there are four nucleotides, one could
start to think that a $4$-adic expansion, which has just four
digits, might be more appropriate. However, note that $4$ is a
composite integer and that such expansion is not suitable since the
corresponding $|\cdot|_4$ absolute value is not a norm but a
pseudonorm and it makes a problem with uniqueness of the distance
between two points. To illustrate this problem let us consider, for
instance, a distance between numbers $4$ and $0$. Then we have $d_4
(0,4) = |4|_4 = \frac{1}{4}$, but on the other hand $d_4 (0,4) =
|2|_4 \, |2|_4 = 1$.

 Recall that there are generally $5$ digits (0,\, 1,\, 2,\, 3,\, 4) in
representation of  $5$-adic numbers. In this approach, we omitted
the digit $0$ to represent a nucleotide, because its consistent
meaning can be only absence of any nucleotide.

Let us note that there are  in general 24  possibilities to connect
four digits with four nucleotides. However, we find that the above
choice seems to be the most appropriate.

An essential property of the $\mathcal{C}_5 (64)$ space of codons is
ultrametric behavior of distances between its elements, which
radically differs from usual distances. One can easily observe that
quadruplets and doublets of codons in the vertebral mitochondrial
code have natural explanation within $5$-adic and $2$-adic
closeness. It follows that degeneracy of the genetic code in the
form of doublets, quadruplets and sextuplets  is direct consequence
of $p$-adic ultrametricity between codons.

There is an important aspect of genetic coding related to particular
connections between codons and amino acids. Namely, which amino acid
corresponds to which multiplet of codons? An answer should be
related to connections between stereochemical  properties of codons
and amino acids.

Let us also note a recent paper \cite{khrennikov2}, where an
ultrametric approach to the genetic code is considered on a diadic
plane.

 \bigskip

\noindent{\bf Acknowledgments}

The work on this paper was partially supported by the Ministry of
Science and Environmental Protection, Serbia, under contract No
144032D. One of the authors (B.D) would like to thank M. Rakocevic
for useful discussions on the genetic code and amino acids.

\end{document}